\documentclass[apl,aip,preprint,amssymb,groupedaddress,floatfix,amsmath,showpacs]{revtex4-1}
\usepackage{graphicx}
\usepackage{bm}
\usepackage[latin1]{inputenc}

\begin{document}

\title{Spin-torque effect on spin wave modes in magnetic nanowires}

\author{Voicu O. Dolocan}\email{voicu.dolocan@im2np.fr} \affiliation{Aix-Marseille University \& IM2NP CNRS, Avenue Escadrille Normandie Niemen, 13397 Marseille, France}

\begin{abstract}
The interaction between a spin polarized dc electrical current and spin wave modes of a cylindrical nanowire is investigated in this report. We found that close to the critical current, the uniform mode is suppressed, while the edge mode starts to propagate into the sample. When the current exceeds the critical value, this phenomenon is even more accentuated. The edge mode becomes the uniform mode of the nanowire. The higher spin wave modes are slowly pushed away by the current until the propagating mode remains.
\end{abstract}

\date{\today}
\maketitle

Reducing the size of materials to the nanoscale is the goal for both fundamental research and technological applications. The underlying physics is different from bulk materials and new functionalities can be realized. Hence, spin waves in confined magnetic nanostructures as nanowires, nanostrips and nanodots were intensively studied both experimentally and theoretically\cite{Hillebrandsbook,Demokritovbook}. Spin wave quantization and localization were observed due to the confinement of finite-size structure. Furthermore, the spectrum of spin waves can be custom shaped by patterning magnetic materials, with the aim of creating magnonic devices in which the information is transported and processed by spin waves\cite{Kruglyak}. Therefore, the understanding of the spin wave spectrum and of their spatial distributions when submitted to external driving forces as the spin transfer torque\cite{Berger,Slonczewski} (STT) are of foremost importance for fundamental and applied research. 

Manipulating the magnetization by a current creates new perspectives for STT-based devices as storage elements or microwave sources. A dc current was shown to excite persistent oscillations of the magnetization\cite{Kiselev} (auto-oscillating mode) at gigahertz frequencies paving the way for creating current-controlled microwave oscillators. Spin wave amplification by a dc current was proposed in magnetic wires for large nonadiabatic parameter\cite{Seo}.

In this letter, we report on the interaction between the spin wave eigenmodes of cylindrical nanowires and an electrical current using micromagnetic simulations. We found that the edge mode, which always exists in finite size wires at the two ends, propagates into the sample due to this interaction while the other spin wave modes are suppressed. The effect is even more significant when the applied current reaches a threshold value and above. The edge mode becomes the uniform mode of the cylinder. As the edge mode frequency has a linear magnetic field dependence, a spin wave with a desired frequency can be chosen in this way.

\begin{figure}[b!]
  \includegraphics[width=7cm]{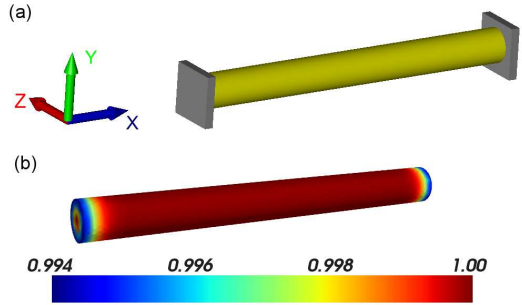}\\
 \caption{\label{Fig.1} (Color online)  (a) Schema of the nanowire connected to two electrods as used in the simulations; (b) Snapshot of the normalized average magnetization along x-axis in the equilibrium configuration. The cylinder has a diameter-to-length ratio of 0.1 with L=300nm.}
\end{figure}

The numerical simulation was conducted with the Nmag package\cite{Fischbacher} using a cylinder with the aspect ratio d/L=0.1, where d is the diameter and the length L is 300nm(Fig.\ref{Fig.1}). The cylinder was discretized into a mesh with a cell size of 4.8nm on the order of the exchange length. The cells are tetrahedral and the cylinder is represented by 15732cells. The material parameters are those of Ni: $\gamma$=188.5GHz/T (g factor of 2.15) and 4$\pi$M$_{s}$=0.6T\cite{Ebels}. In all the simulations, the magnetization was first relaxed to equilibrium applying a large magnetic field in the long axis direction ($x$-axis) using a large damping parameter $\alpha$=0.5. For the subsequent dynamical simulations, the damping parameter was chosen as 0.015.

\begin{figure}[t!]
  \includegraphics[width=6cm]{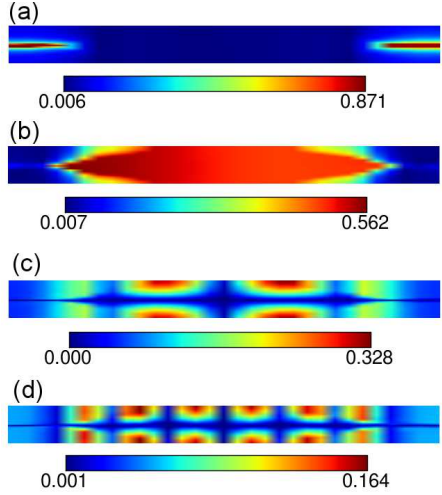}\\
 \caption{\label{Fig.2} (Color online)  Spin wave eigenmodes in a nano-cylinder excited by magnetic fields: (a) edge mode, (b) fundamental mode, (c) and (d) mixed modes. The power FFT distributions of dynamical magnetization m$_z$ are shown in a false color code that is normalized to 1. The external field is applied along the long axis.}
\end{figure}

We begin by showing (Fig.\ref{Fig.2}) the spin wave modes that appear in cylindrical nanowires of reduced size under the influence of magnetic fields. The external magnetic field is applied along the cylinder axis. Although we use full 3D micromagnetic simulations, it is easier to analyze the spatial distribution of the modes in 2D. Hence, only the variation of the dynamical magnetization in a central plane is displayed. In the shown power Fast Fourier Transform (FFT) images, the variable magnetization is normalized to 1 for each applied configuration (each value of applied current and magnetic field). Two major peaks appear in the power spectrum of the variable magnetization m$_z$: one corresponding to the uniform mode (panel (b)) which is the largest peak, and one corresponding to the edge mode (panel (a)) which is localized at the two ends of the cylinder. These modes always appear in finite size nanowires\cite{Dolocan}. Other spin wave modes appear at higher frequencies, which we termed as mixed modes, and correspond to modes with nodes along the short and long axes of the cylinder. Modes with similar characteristics were calculated and detected in elliptical dots\cite{Gubbiotti}. The mixed modes can be readily identified considering that the dynamic magnetization has a harmonic dependence on the $x$ axis and the spin wave vector is quantized. 

\begin{figure}[t!]
  \includegraphics[width=6cm]{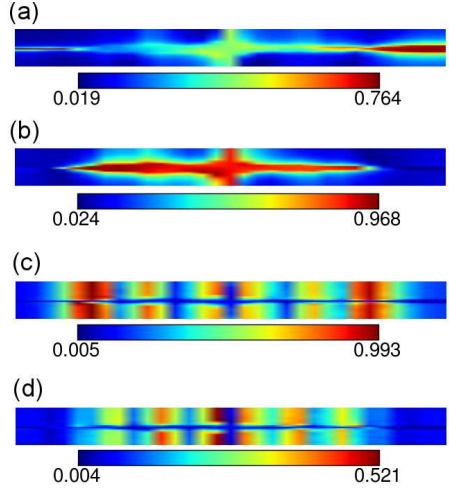}\\
 \caption{\label{Fig.3} (Color online) Interaction of the spin wave modes in a nano-cylinder with a dc current. The same spin wave-modes as in Fig.\ref{Fig.2} are shown.  The power FFT distributions of dynamical magnetization m$_z$ are shown in a false color code that is normalized to 1.}
\end{figure}

The changes in mode profiles, when a dc current is applied along the principal axis of the cylinder, are shown in Fig.\ref{Fig.3}. These profiles are calculated as follows: the same modes as in Fig.\ref{Fig.2} are excited, then the external magnetic field is switched off and only a dc current of amplitude 10$^{11}$A/m$^2$ is applied along the $x$ axis for 5ns. The amplitude of the current is chosen below the critical value. The influence of the dc current is threefold: it modifies the frequencies of the modes, it changes the spatial profile of the modes and excites a hybrid mode between the edge and the fundamental mode.


\begin{figure*}[t!]
  \includegraphics[width=13cm]{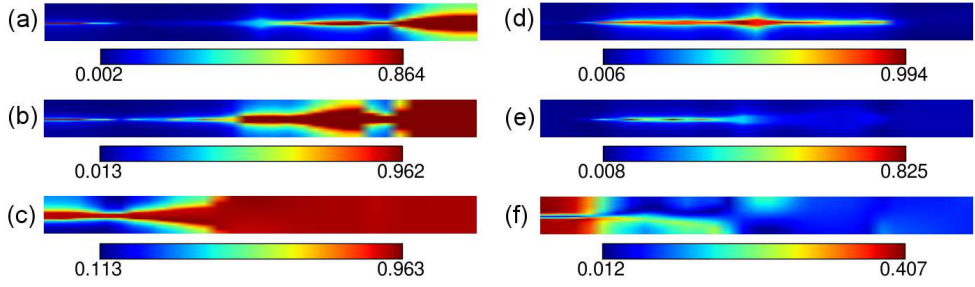}\\
 \caption{\label{Fig.4} (Color online)  Interaction of the edge mode (left) and uniform mode (right) with a dc current of amplitude: 3$\times 10^{11}$A/m$^2$ in (a) and (d), 5$\times 10^{11}$A/m$^2$ in (b) and (e), 7.5$\times 10^{11}$A/m$^2$ in (c) and (f). Above the critical current, the edge mode propagates in the cylinder while the uniform mode is suppressed. The current was applied for 5ns and no domain walls are present. The power FFT distributions of dynamical magnetization m$_z$ are shown in a false color code that is normalized to 1.}
\end{figure*}

To understand the interaction of the current with the variable magnetization, we analyze the Landau-Lifschitz-Gilbert (LLG) equation including damping and STT terms\cite{ZhangLi} as used in the numerical simulation:

\begin{align}
\label{eq1}
\frac{\partial\textbf{m}}{\partial t}(\textbf{r},t)  = & -\gamma\textbf{m}\times\textbf{H}_{eff} + \alpha \textbf{m}\times\frac{\partial \textbf{m}}{\partial t} \nonumber \\ 
& + (\textbf{u} \cdot \nabla)\textbf{m} - \beta(\textbf{m}\times(\textbf{u}\cdot\nabla)\textbf{m})
\end{align}

\noindent where $\textbf{m}=\frac{\textbf{M}}{M_S}$ is the unit vector along the local magnetization, $\gamma$ is the gyromagnetic ratio, $\textbf{H}_{eff}$ is the effective field inside the sample that includes the external field $\textbf{H}_{ext}$, the exchange field and the demagnetizing field (excluding crystal anisotropy), $\alpha$ is the damping parameter and $\beta$ the nonadiabatic parameter. The effect of the spin current is introduced through the spin current drift velocity $\textbf{u}=\frac{jP\mu_B}{eM_S}$ with P the spin polarization of the current, $\mu_B$ the Bohr magneton, $e$ the electron charge and $j$ the current density. We consider that the saturation magnetization M$_{s}$ is uniform in the sample and substantially larger than the variable magnetization. The effect of Oersted field is not taken into account, as for small nanostructures its effect is small compared to the spin-torque effect.

To find the normal modes of the dynamic magnetization we search for non zero solutions of the linearized LLG equation. Analytically, the Eq. \eqref{eq1} can be integrated in only a small number of cases and usually approximations are made on the scalar potential or the magnetization. Without the current, the mode profiles can be readily explained assuming the dynamical magnetization of the form $m \sim J_n(qr)\cos(\omega t - k_x^l x)$\cite{Dolocan} with $J_n(qr)$ the Bessel function of order n, $q$ and $k_x^l$ the radial and longitudinal wave numbers respectively. Here $l$ indexes the longitudinal modes and $n$ the radial ones ($ln$ modes) and we also assumed that the 3D profile of the variable magnetization can be expressed as a product of eigenfunctions for longitudinal and radial directions. This expression is valid in the effective length, between the turning surfaces where stationary waves are formed\cite{Demokritovbook}. In this way, the mixed spin wave modes in Fig.\ref{Fig.2} (c) and (d) can be identified as (2,1) mode and respectively the (6,1) mode.

The analytic solutions of the Eq.\eqref{eq1} with STT terms are even more complicated. Assuming, for simplicity, only a harmonic longitudinal dependence ($m\sim\exp(i\omega t-k_x x)$) of the dynamical magnetization and that the dc current is applied along the $x$ axis, the solutions for an infinite cylinder are of the from\cite{He}:

\begin{equation}
\label{eq2}
\omega(k) = \frac{k_x u(1+\alpha\beta)\pm\gamma H_{eff} + i[\alpha\gamma H_{eff} \pm k_x u(\alpha-\beta)]}{1+\alpha^2}
\end{equation}

The minimum critical current can be determined at the onset of the instability when $Im[\omega]=0$:

\begin{equation}
\label{eq3}
j_c = \sqrt{\frac{2A}{M_s}}\frac{eM_s\gamma\sqrt{H_{eff}}}{P\mu_B\sqrt{\vert\frac{\beta}{\alpha}-1\vert}}
\end{equation}

The dispersion relation becomes in this case  $\omega(k) \simeq k_x u+\gamma H_{eff}$, which shows how the current influences the spin wave spectrum. The modification of the spin waves by the current was shown to be a general case for ferromagnets, with the magnon energy modified by a quantity proportional to $\textbf{j}\cdot\textbf{k}$\cite{Fernandez}. This explains the spin-wave Doppler shift, but not the spatial distribution of the spin-waves. The general analytical solutions of the LLG equation considering the boundary conditions are difficult to calculate and interpret physically. We do not take into account in the analytical model the spin-wave attenuation\cite{Seo}, because in our case it has a small influence on the dispersion relation. Our nanowire is of finite aspect ratio with short length (300nm), while the attenuation length of excited waves is larger. The phenomenon presented here, the propagation of the edge mode is not influenced by the attenuation length or the $\beta$ parameter. 

From Eq.\eqref{eq3}, using the parameters of our numerical simulation $P=0.7$, $\alpha=0.015$, $\beta=0.05$ and $A=10.5$pJ/m, we obtain $j_c=3\times10^{11}$A/m$^2$. This value\cite{comment} agrees well with the one obtained from micromagnetic simulations at the onset of the instability. Increasing the current above it will nucleate a domain wall(not shown). Although the value of the applied current in Fig.\ref{Fig.3} was just below the critical value, the effect of the current is larger on the uniform mode. Its spatial profile changes greatly, partly to the Gilbert damping (through $\alpha$) and partly to the spin torque effect. The STT acts like a positive damping for the uniform mode, contrary to the edge mode which appears enhanced (negative damping). This effect becomes even larger when the current is increased to the critical value and beyond. 

In Fig.\ref{Fig.4}, we present the central result of this letter. We observe that with increasing dc current, close to the threshold value and above, the edge mode propagates into the sample (panels (a)-(c)). For the panels (b) et (c), which correspond to magnitudes above threshold, the current was applied for 5ns and then stopped to avoid the formation of the domain wall (current pulses). No domain wall is present for these cases, the propagation shown is of the spin wave edge mode alone which has almost the same frequency in panels (a) to (c). After the current is stopped, the spin wave continues to propagate (images not shown) and it becomes the uniform mode of the cylinder while the amplitude of the variable magnetization decreases exponentially in time. The edge mode frequency does not seem to be much affected by the current, its variation between the three presented cases is around the spacing in the frequency domain of the raw data (0.1GHz). The previous uniform mode is highly damped until it disappears as can be observed in panels (d) to (f) and the main peak in the power spectrum becomes the one from the edge mode. This mode becomes the uniform mode of the nanowire.


\begin{figure}[b!]
  \includegraphics[width=7cm]{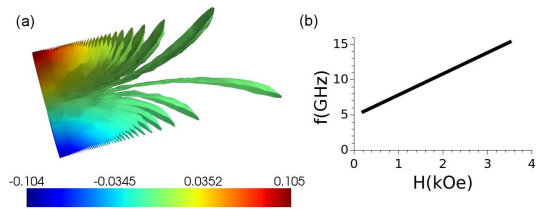}\\
 \caption{\label{Fig.5} (Color online) (a) Isosurfaces of the variable magnetization at one edge of the nanowire. The local gradient of $\textbf{m}$ is normal to the isosurfaces; (b) Variation of the edge mode frequency with applied magnetic field in absence of current.}
\end{figure}

The propagation of the edge mode can be traced back to the spin torque terms in the LLG equation. The STT relates the time gradient of the variable magnetization to its space gradient. Its main action is there where a spatial gradient exists: close to the edges (see Fig.\ref{Fig.1}). In Fig.\ref{Fig.5}(a), the calculated isosurfaces of $\textbf{m}$ are presented at one edge of the nanowire, before the polarized current was applied. The local spatial gradient is normal to the isosurfaces and has the direction of increasing $\textbf{m}$. The first interaction with the STT occurs here and the mode is pushed into the cylinder if the current is large enough. The frequency of the edge mode varies linearly with applied magnetic field in absence of the current, as shown in Fig.\ref{Fig.5}(b)\cite{Dolocan}. The influence of the current on the edge mode frequency seems minor and is proportional to $k_xu$ in the simple analytical model. For a current density of 10$^{11}$A/m$^2$, $u=8.5$m/s therefore the effect of the current is small as the wavelength of the edge mode is large.  

The propagation of the edge mode also occurs when a magnetic field is present (not shown). For fields as larges as 0.3kOe applied along +$x$ axis, we observed the propagation of the edge mode into the sample when the current exceeds the critical value while the uniform mode becomes suppressed. However, the propagation of the edge mode is slower as it acts against the applied magnetic field. The edge mode frequency and spatial distribution is field dependent, the mode being localized closer to the edge with increasing magnetic field (without an applied current). This localization doesn't seem to affect the propagation of the mode when a current is applied. The localization of the edge mode was also observed experimentally in microscopic ellipses\cite{Jersch}.


\begin{figure}[t!]
  \includegraphics[width=6cm]{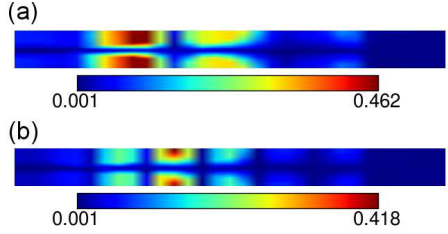}\\
 \caption{\label{Fig.6} (Color online) Effect of the spin torque on the spin wave modes (2,1) and (6,1) after a dc current of 5$\times10^{11}$A/m$^2$ is applied for 5ns. The profile of the propagating edge mode is shown in Fig.\ref{Fig.4}(b). The power FFT distributions of dynamical magnetization m$_z$ are shown in a false color code that is normalized to 1.}
\end{figure}

The effect of the dc current on the spin wave modes (2,1) and (6,1) is shown in Fig.\ref{Fig.6}. We observe that the spin waves are pushed away by the propagating mode while the amplitude of the modes decreases. The displayed configuration is the same as in the Fig.\ref{Fig.4} (b) and (e). The spatial profile of the edge mode corresponds with the suppressing of the modes on the right side. In our calculation, only the propagating mode remains after 10ns, all the other modes being suppressed.

The interaction between the spin waves and the dc current was calculated assuming that the STT terms (additional terms in the LLG equation) act entirely to rotate the magnetization (in the direction of the moment of incoming electrons) and not to excite other spin wave modes as also proposed\cite{Tsoi,Bazaliy}. In real materials, other sources of damping can occur which can lead to the violation of the momentum conservation law\cite{Gurevich}. This allows for the coupling of different spin waves (two magnon scattering, etc.) which can modify the spectrum of spin waves in the presence of the current\cite{Fernandez}. Quantitatively, the exact damping is difficult to estimate. In our ideal nanowire, the damping parameter $\alpha$ is assumed constant and independent of current and magnetic field. It was proposed that the LLG equation should be modified for micromagnetic simulations, to account for variable damping\cite{Zhang}. However, the modification will be important for the case of rapidly variation of magnetization as for short wavelength spin waves and narrow domain walls and will not affect the propagation of the edge mode in our case.

In summary, we studied the behavior of spin waves in the presence of an electrical current for a cylindrical nanowire of reduced aspect ratio. We showed that the current modifies the spin-wave spectrum and the spatial distribution of the modes considerably near the threshold current. The edge mode is pushed inside the nanowire while the other modes are slowly suppressed. Above the threshold, this effect is greatly enhanced. This effect could be experimentally detected by time resolved spin wave spectroscopy\cite{Covington}. Nowadays, the fabrication of Ni nanowires of 30nm diameter by electrodeposition is facile and cost effective\cite{Yoo}. Large arrays of nanowires equally spaced can be easily fabricated and used as waveguides for the propagation of spin waves. As the edge mode has a frequency that varies linearly with the magnetic field, a spin wave with a chosen frequency can be generated and sent along the magnetic waveguide (nanowire). This could be the basis for creating an active microwave device.


The author wish to thank L. Raymond for the access to the Merlin cluster and is grateful for the support of the NANOMAG platform by FEDER.


\end{document}